\documentclass[%
reprint,
superscriptaddress,
amsmath,
amsymb,
aps,
prb,
showpacs,
floatfix,
]{revtex4-2}

\usepackage{graphicx}
\usepackage{dcolumn}
\usepackage{bm}
\usepackage{comment}
\usepackage{hyperref}
\begin{document}

\preprint{APS/123-QED}

\title{Chaos-Assisted Dynamical Tunneling in Flat Band Superwires}

\author{Anton M.~Graf}
\affiliation{Harvard John A. Paulson School of Engineering and Applied Sciences, Harvard University, Cambridge, MA 02138, USA}
\affiliation{Department of Physics, Harvard University, Cambridge, MA 02138, USA}
\affiliation{Department of Chemistry and Chemical Biology, Harvard University, Cambridge, Massachusetts 02138, USA}

\author{Ke Lin}
\affiliation{Department of Physics, Harvard University, Cambridge, MA 02138, USA}
\affiliation{Zhiyuan College, Shanghai Jiao Tong University, Shanghai, Shanghai 200240, China}

\author{MyeongSeo~Kim}
\affiliation{Department of Physics, Harvard University, Cambridge, MA 02138, USA}
\affiliation{Harvard College, Harvard University, Cambridge, MA 02138, USA}

\author{Joonas~Keski-Rahkonen}
\affiliation{Department of Physics, Harvard University, Cambridge, MA 02138, USA}
\affiliation{Department of Chemistry and Chemical Biology, Harvard University, Cambridge, Massachusetts 02138, USA}

\author{Alvar Daza}
\affiliation{Department of Chemistry and Chemical Biology, Harvard University, Cambridge, Massachusetts 02138, USA}
\affiliation{Nonlinear Dynamics, Chaos and Complex Systems Group, Departamento de  F\'isica, Universidad Rey Juan Carlos,Tulip\'an s/n, 28933 M\'ostoles, Madrid}

\author{E.J.~Heller}
\affiliation{Department of Physics, Harvard University, Cambridge, MA 02138, USA}
\affiliation{Department of Chemistry and Chemical Biology, Harvard University, Cambridge, Massachusetts 02138, USA}


\date{\today}

\begin{abstract}

Recent theoretical investigations have revealed  unconventional transport mechanisms within high Brilliouin zones of two-dimensional superlattices. Electrons can navigate along channels we call superwires,  gently guided without brute force confinement. Such dynamical confinement is caused by weak superlattice deflections, markedly different from the static or energetic confinement observed in traditional wave guides or one-dimensional electron wires. The quantum properties of superwires give rise to elastic dynamical tunneling,  linking disjoint regions of the corresponding classical  phase space, and enabling the emergence of several parallel channels. This paper provides the underlying theory and mechanisms that facilitate dynamical tunneling assisted by chaos in periodic lattices. Moreover, we show that the mechanism of dynamical tunneling can be effectively conceptualized through the lens of a paraxial approximation. Our results further reveal that superwires predominantly exist within flat bands, emerging from eigenstates that represent linear combinations of conventional degenerate Bloch states. Finally, we quantify tunneling rates across various lattice configurations, and demonstrate the tunneling can be suppressed in a controlled fashion, illustrating potential implications in future nanodevices. 

\end{abstract}

\maketitle

\section{Introduction}

\noindent Two-dimensional (2D) materials and electronic systems have become a focal point in solid-state physics research, offering numerous potential applications in the realm of electronics.~\cite{zeng2021} Motivated by the discovery of twisted bilayer graphene and advancements in artificial superlattices (see, e.g. Refs.~\cite{lu2013twisting, andrei2020graphene,Rasanen2012,Paavilainen2016,Polini2013}), various versions of \emph{soft} Lorentz gases, i.e. arrays of fixed scatterers, have been explored, including trigonometric functions~\cite{lorentzcos1}, Lennard-Jones potentials~\cite{lorentzlennard}, and hard-wall disks with reduced potentials~\cite{lorentzlowpot}. In fact, a recent study has unveiled a complex interaction between trapped and ballistic periodic orbits, leading to both normal and anomalous diffusion in a soft Lorentz gas model that replaces the disks with Fermi-type potential profile~\cite{lorentzgas}. Moreover, recent work by some of the present authors revealed  branched flow~\cite{heller2021branched} in a soft Lorentz gas, thus not restricted to disordered media wherein a travelling quantum wave scattered by random smooth obstacles generates a tree-like ``branched flow'' pattern~\cite{doi:10.1073/pnas.2110285118}. 

In addition to the branched flow, indefinitely stable branches were also seen to form in the channels of a periodic lattice of smooth bumps or dips~\cite{doi:10.1073/pnas.2110285118}. Within these stable branches, propagating waves are dynamically confined to quasi one-dimensional channels called \emph{superwires}. In contrast to conventional wires relying on deep potential barriers, the traveling waves in these superwires have enough energy to surmount the channel potential and thus would not be restricted into the channel. The formation of the superwires is governed by the underlying dynamics of the system, a mechanism of confinement that diverges fundamentally from the likes of Luttinger liquids or from states induced by static disorder, such as those seen in Anderson localization. This introduces a distinctive framework, marking a clear departure from traditional paradigms of wave confinement.

In this work we show that for identical parallel wires tunneling between them inevitably occurs, and flux slowly flows to neighboring channels.  The slow tunneling rate gives rise to  nearly flat bands in the direction  perpendicular to the wire propagation. Since the classical motion is stable flow down the channel,  with no insurmountable potential barrier, and chaos lurks in parts of phase space, we will attribute this leakage to side channels as "chaos-assisted dynamical tunneling”~\cite{davis1981quantum, dyntun,PhysRevE.50.145}. In this work, we explore the mechanisms behind dynamical tunneling in superwires and quantify tunneling rates across different lattice configurations. We can understand much of the physics through a simple model based on the paraxial approximation.  We can selectively suppress or enhance tunneling between channels in a controlled manner, offering a pathway to leverage superwires in advanced nanodevices of the future, as in next-generation high-precision measurement and sensing instruments.

We begin in Section \ref{Section2} with examples of superwire flow over many lattice cells. Subsection \ref{Subsec1} discusses parameters necessary for highly stable superwires. Subsection \ref{Subsec2} describes the flat bands associated with superwires, and discusses the construction of superwire states using degenerate Bloch waves. Subsection \ref{Subsec3} shows the generation of superwire eigenstates through quantum wavepacket dynamics simulations. Subsection \ref{Subsec4} aims to simplify the concept of dynamical tunneling, presenting it through the lens of a paraxial approximation to make the quantum mechanical principles more accessible. Section \ref{Section3} focuses on comparing the quantum and classical aspects of superwires. Here we justify the dynamical nature of tunneling, using classical simulations and Poincaré surfaces of section (PSS) as evidence. In Section \ref{Section4}, we investigate how slight alterations in the geometry of superlattice features can influence dynamical tunneling. Our analysis, which quantifies tunneling rates in different lattice channels, reveals that even minor changes can significantly suppress dynamical tunneling. Section \ref{Section5} concludes the paper by summarizing our key results and proposing directions for future investigation.

\section{Quantum Superwires, Dynamical Tunneling, and Flat Bands} \label{Section2}

\begin{figure*}[ht]
\includegraphics[width=1\linewidth]{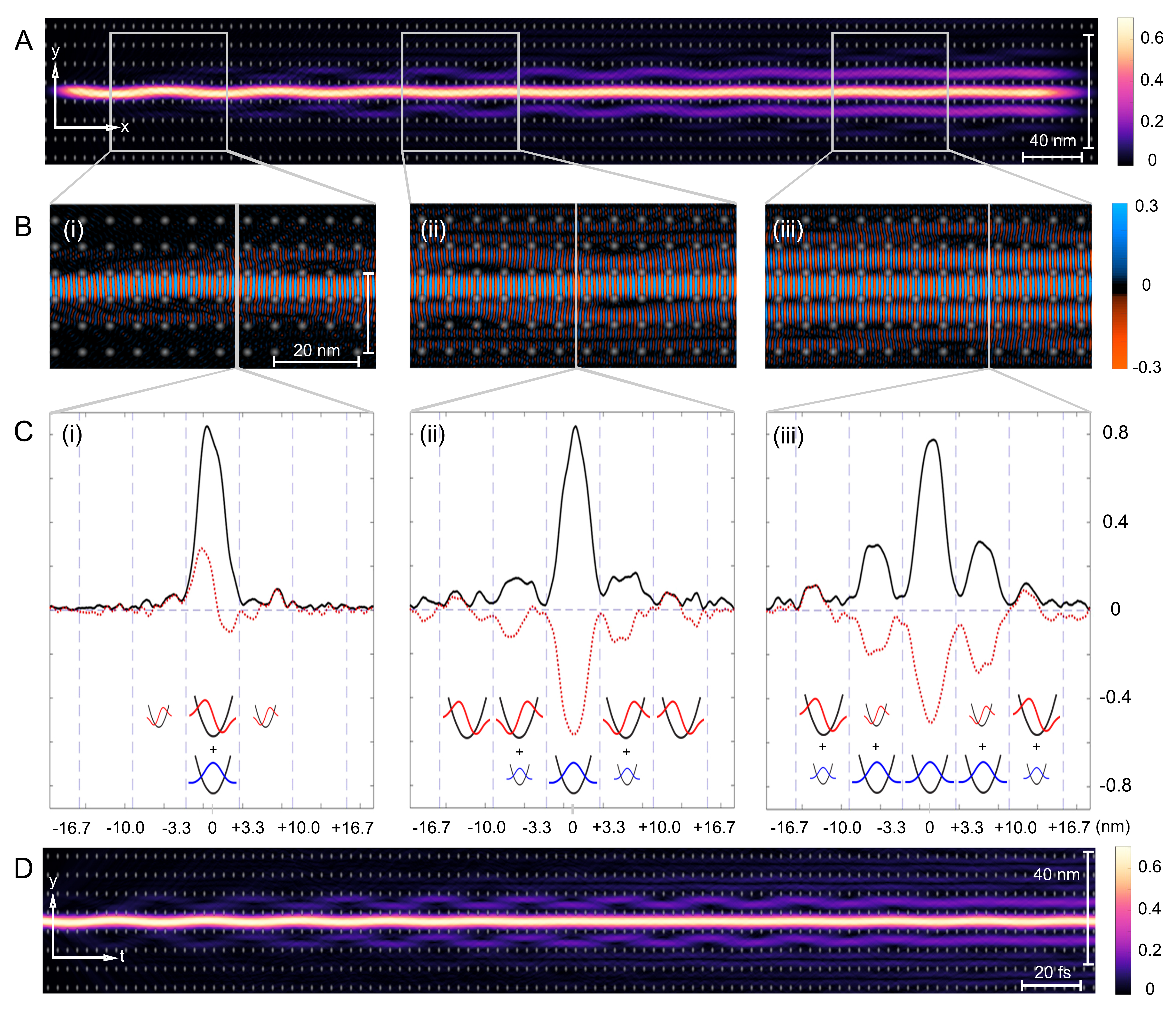}
\caption{Chaos-assisted dynamical Tunneling in superwires. A) The probability density $P(x,y)$ of a superwire eigenstate extracted from quantum dynamics simulation. B) Real part of the wavefunction $\Re(\psi(y,t))$ magnified in the regions (i), (ii), and (iii) indicated by grey boxes. C)  Cross-section $\Re(\psi(x,y))$ (red dotted line) and $|\psi(x,y)|$ (black line) at $x=90.8$~nm, $x=295.7$~nm, and $x=598.5$~nm (indicated by grey lines for (i), (ii), and (iii) in panels B). The oscillator eigenstates at the bottom schematically explain the contributions of the ground state (GS) in blue and first excited state components (1ES) in red in each channel. Large oscillator symbols indicate major contribution, small symbols a minor, and missing symbols negligible contribution to the wavefunction.  D) $P(x,y)$ of a superwire in time   extracted from a quasi one-dimensional model \textit{via} paraxial approximation.}
\label{fig1}
\end{figure*}

In a nutshell, superwires give rise to electron transport in solids confined to narrow channels, almost one-dimensional paths. This behavior emerges a consequence of quasi-periodic deflections from the static, non-integrable superlattice structure accommodating multiple electron wavelengths between its features, corresponding to higher Brillouin zones.  The energy of the electron wave significantly exceeds the peak and trough values of the superlattice potential, enabling them to surmount potential barriers while still being confined to channels. Either the Moiré patterns of twisted atomic layers or direct fabrication can achieve this regime. As  discussed above, we probe the dynamics of superwires within a two-dimensional soft Lorentz gas, which were recently reported in Ref.~\cite{doi:10.1073/pnas.2110285118}. 
More specifically, we aim to understand the quantum nature of superwires within superlattices and investigate the fundamental mechanisms behind their robust dynamical properties, particularly in relation to chaos-assisted dynamical tunneling. Figure~\ref{fig1} collects our main results: A superwire eigenstate illustrating the dynamics of an electronic wavepacket propagating along a single channel but eventually leaking into adjacent channels, a feature which can be understood in terms of a simple paraxial approximation. The remainder of the paper explores and  analyzes these results.

\subsection{Setup of superwire systems} \label{Subsec1}

It was previously demonstrated that superwires appear across a diverse range of superlattice geometries, with numerical evidence for both square and trigonal lattices formed by concave (dips) or convex (bumps) potential features~\cite{doi:10.1073/pnas.2110285118}. In the same spirit, we study a simple example model, characterized by smooth potential bumps at equidistant lattice points in a square lattice, with five to six electron wavelengths per unit cell. The potential is specified by Fermi-type functions centered at each lattice point~\cite{lorentzgas, doi:10.1073/pnas.2110285118}.

\begin{figure}[h!]
    \centering
    \includegraphics[width=0.95\linewidth]{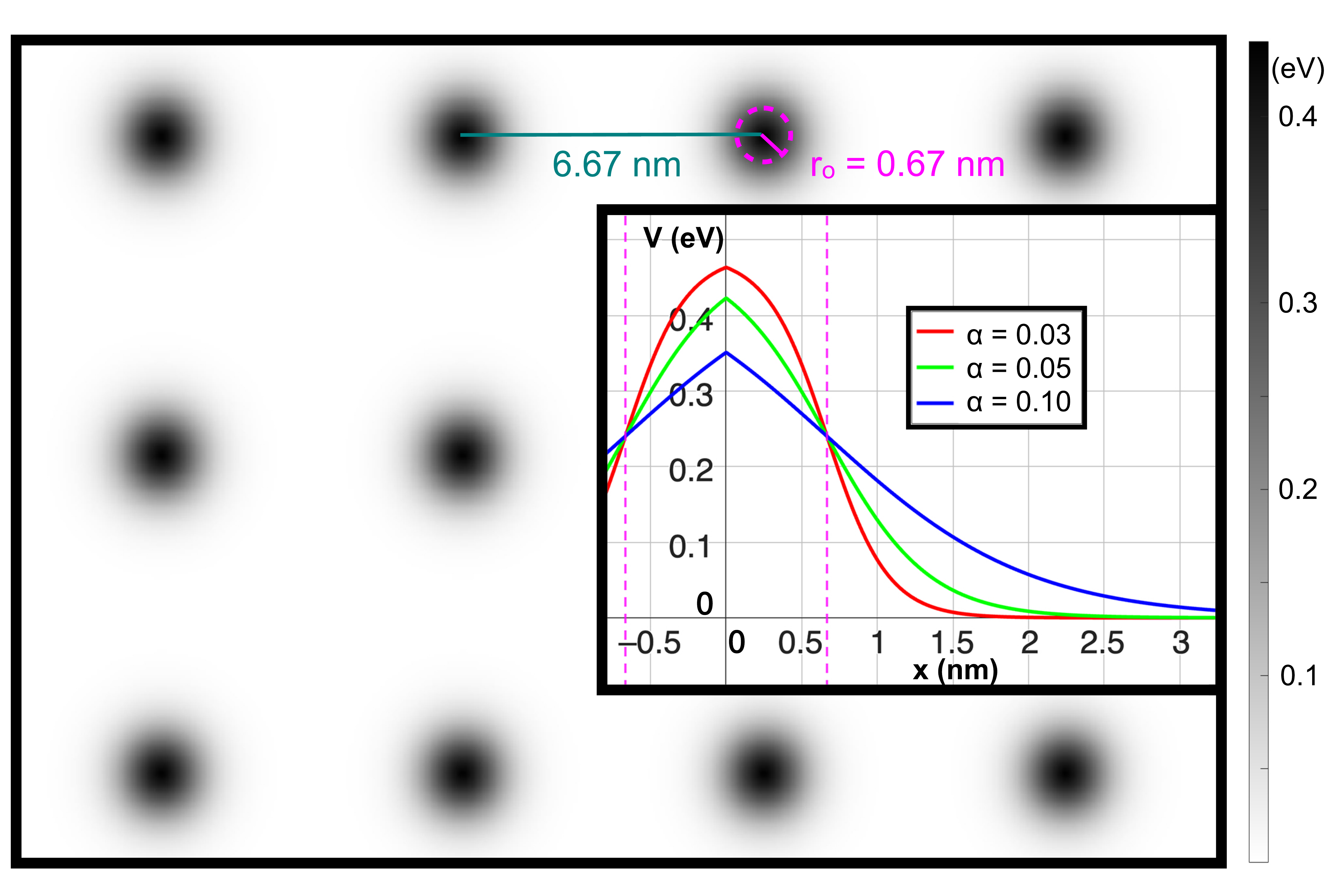}
\caption{Superlattice structure. The potential is depicted through grayscale dots generated by Fermi functions. The inset presents a cross-sectional view through the center of the bump, highlighting the smoothness and relative height of the bumps for three different $\alpha$ values. The pink dashed line represents the (effective) radius, and the dark green line denotes the lattice constant chosen for the simulations.}
\label{fig4}
\end{figure}

More specifically, we consider the following Hamiltonian $H(\textbf{r},\textbf{k})$ 
\begin{equation}
H(\mathbf{r},\mathbf{k})=\frac{\hbar^2\vert \mathbf{k} \vert^2}{2m}+
\sum_{\mathbf{r}_i \in S} \frac{V_0}{1+\exp[(|\mathbf{r} -\mathbf{r}_i \vert - r_0)/\alpha]}, 
\label{Ham}
\end{equation}
which yields a mixed system containing features of both non-integrable (chaotic) and integrable (regular) dynamics in classical and quantum considerations. In the Hamiltonian, $\textbf{r}$ and $\textbf{k}$ are the location and momentum of an electron with mass $m$, respectively. The potential is composed of individual bumps located at a (finite) set of lattice points $\mathbf{r}_i $ that are characterized by their amplitude $V_0$ and (effective) radius $r_0$. The smoothness of the Fermi-potential bumps is controlled by $\alpha$. We want to emphasize that this potential model potentials is feasible for artificial superlattice experiments~\cite{lu2013twisting, andrei2020graphene,Rasanen2012,Paavilainen2016,Polini2013}). Without losing generality, we set
the distance between bums to be 0.4 nm, and  $\alpha$  to be 0.2 nm, with a lattice spacing of \textit{ca} 6.67 nm. Remarkably, the potential features do not need to be strong for superwires to form. The height of the potential is scaled to approximately 0.45 eV at the top and 0 eV at the bottom. The resulting superlattice structure is illustrated in Figure~\ref{fig4}.

\subsection{Band structure} \label{Subsec2}

Analyzing the band structure of the system offers additional insights with experimental significance for the study of quantum transport superlattices. High index bands, are  crucial for comprehending superwire physics. We focus on the superlattice potential, rather than the atomic lattice, simplifying the investigation of the band structure in high Brillouin zones of a 2D square lattice. The procedure is explained in \ref{A2}. 

\begin{figure*}[ht]
\includegraphics[width=1\linewidth]{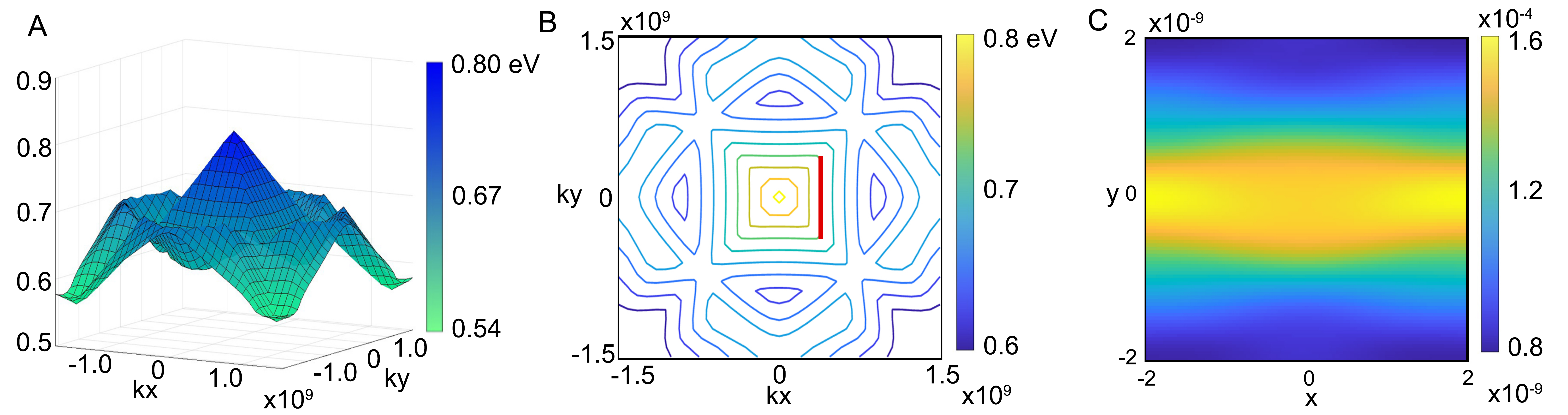}
\caption{The band structure $B(k_x,k_y,E)$  of the $10^\text{th}$ band where superwire states are found. The band shows features of a four-sided pyramid with flat bands along the side. B) Contour map of the $10^\text{th}$ band $B(k_x,k_y,E)$ emphasizing the flat bands. The red line indicates the flat band with degenerate Bloch states that are used to generate the eigenstate shown on the right. C) Probability density of Superwire eigenstate that has maximal overlap with the degenerate Bloch states on the red flat band contour.}
\label{fig2}
\end{figure*}

Figure \ref{fig2}A presents the projected band structure, $B(k_x, k_y, E)$, of the $10^\text{th}$ band, which is identified as one of the bands exhibiting superwire states within the superlattices discussed in this study.
\twocolumngrid\
One of the bands we have found has a geometry close to a four-sided pyramid with a peak at the center. There is a gap between the peak of the pyramid and the  band above. The contour map in panel B reveals the existence of \emph{flat bands} along the sides of the pyramidal structure.  The flat bands shown are uniform along one momentum component; at the corners they turn between flat in $k_x$ to flat in $k_y$. Flat bands are of considerable interest in the study of engineered superlattice structures~\cite{li2021imaging, leykam2018artificial, tian2023two, tamura2002flat, mao2020evidence, PhysRevB.108.235418} and superconductivity~\cite{li2020artificial, PhysRevB.107.214508, tian2023evidence, balents2020superconductivity}.
Flat bands play an important  role in twisted bilayer graphene superconductivity~\cite{lisi2021observation, yankowitz2019tuning, marchenko2018extremely, oh2021evidence,chebrolu2019flat}, and their presence is vital for the dynamics of superwires.
Energy eigenstates can be  described by Bloch waves which have a given pseudomomentum and energy. Our superwire eigenstates, shown in Figure~\ref{fig1}A for example, are analogous to a focused laser beam in free space: The focused laser beam has a single frequency (fixed energy), but many plane waves of different wave vector (different momentum) are required to make it focus.  Here we have similarly sacrificed pseudomomentum yet have maintained fixed energy.  The beams are made of linear combination of Bloch waves with momentum along a flat band contour, as highlighted by the red line. Using the method of generating functions, as detailed in Ref.~\cite{heller2018semiclassical}, we determine the overlaps between the superwire state and all states along the contour marked in red. The probability distribution $P(x,y)$ of the eigenstate is seen in panel C, producing a superwire eigenstate.

\subsection{Wavepacket Dynamics and Eigenstates} \label{Subsec3}

Our investigation of dynamical tunneling starts with a squeezed Gaussian that is a common tool in multiple fields, such as quantum optics~\cite{scully1997quantum, walls2007quantum} and quantum scarring~\cite{heller2018semiclassical, PhysRevB.96.094204.2017, PhysRevLett.123.214101.2019}, launched along the superwire with high fidelity. The initial wavepacket is elongated along the center channel in x,  and chosen to have a large overlap with superwire eigenstates: 
\begin{equation}
\begin{aligned}
\psi(0) = \exp \Bigg[&-\left(\frac{(x-x_0) \cos(\theta) - (y-y_0) \sin(\theta)}{\sqrt{\sigma_x}}\right)^2 \\
&-\left(\frac{(x-x_0) \sin(\theta) + (y-y_0) \cos(\theta)}{\sqrt{\sigma_y}}\right)^2 \\
&+ i(k_x x + k_y y) \Bigg].
\end{aligned}
\end{equation}
In order to study the tunneling into neighboring channels, we isolate the initial flow into one channel with high fidelity $>99\%$ explained in Section \ref{Subsec3}. Therefore, we choose standard deviations of $\sigma_x=10\,\textrm{nm}$ and $\sigma_y=2\,\textrm{nm}$ in the x- and y-directions, respectively, and the momentum $k_x \gg k_y$. As a consequence, the wavepacket predominantly moves in the +x-direction. The local electron wavelength, $\lambda_{e}=1.31\,\textrm{nm}$, corresponds to roughly $5\lambda_{e}$ per unit cell along the propagation direction. Furthermore, we initialized the wavepacket with momentum directed towards the right,  targeting the center channel.
We employ the third-order split-operator method \ref{A2} to propagate the wavepacket according to the time-dependent Schrödinger equation
\begin{equation}
\hat{H}\psi(x, y, t) = i\hbar\frac{\partial}{\partial, t}\psi(x,y, t)
\end{equation}
where the Hamiltonian is given by equation~\ref{Ham}. Throughout the simulation, we find that the probability density   propagates from left to right, while remaining almost exclusively confined to the central channel. As the wavefunction evolves, it reaches the rightmost boundary of the position grid, spanning several unit cells of the superlattice potential. To enable the computation of a large grid where tunneling can be studied we used a matrix extension technique, with further details provided in \ref{A1}. 
In supplement to analysis of the wavepacket evolution, we construct an eigenstate of the Hamiltionian from the given wavepacket as 
\begin{equation}
\vert \psi_n(x,y) \rangle  \approx \lim_{T \to \infty} \frac{1}{2 T} \int_{-T}^{T} \exp(-E_n t) \vert \psi(x,y,t) \rangle \, \textrm{d} t, 
\end{equation}
which gives the wanted eigenstate after an convenient renormalization to the simulation box size. The studied energy $E_n$ of the eigenstate is selected as the highest weight in the power spectrum
\begin{equation}
S(E) = \frac{1}{2\pi\hbar} \left \vert \int_{-T}^{T} e^{iEt/\hbar} \langle \psi(0) | \psi(t) \rangle \, dt \right \vert,
\end{equation}
which is the Fourier transform of the autocorrelation function \( c(t) = \langle \psi(0) | \psi(t) \rangle \). 
Figure~\ref{fig1} illustrates the superwire confinement and tunneling mechanism \textit{via} a constructed eigenstates at an energy of 0.9 eV, double the maximum of $V_0$ whose corresponding probability distribution $P(x,y)$ is shown in Figure~\ref{fig1}A. The probability density peaks in the central channel near the left launch point. As the wavepacket propagates to the right, it expands into adjacent channels through dynamical tunneling. Subsequently, a significant portion of density $P(x,y)$ tunnels into neighboring channels after a critical time threshold, defying classical phase space boundaries set by the stability islands. This behavior, elucidated in Section~\ref{Section3} and numerically validated in Figure~\ref{fig3}, results in probability growth in adjacent channels at the expense of the middle channel. In Figure~\ref{fig1}B, the real part of the wavefunction $\Re(\psi)$, is visualized with red representing positive amplitude and blue indicating negative amplitude against a grayscale background delineating lattice potential bumps in white and uniform potential in black. Additionally, Figure~\ref{fig1} C presents a cross-sectional view of the real part $\Re(\psi)$, in red and its magnitude in black across the y-direction, averaged over the examined section, with grey dashed lines demarcating lattice points to facilitate channel separation.

\subsection{Understanding Tunneling through the Lens of a Paraxial Approximation} \label{Subsec4}

The paraxial approximation significantly simplifies the analysis of superwire dynamics, particularly in understanding the mechanism of dynamical tunneling. By applying this approximation under the condition $k_x~\gg~k_y$, and acknowledging the superwire states'  x-direction propagation, we  focus on the y-component of the wavefunction. We substitute the x-direction (distance of the superwire) with a temporal dimension $x \longrightarrow t$. We then solve the time-dependent Schrödinger equation for a quasi-one-dimensional system for a potential that varies periodically in time. The motion in the x-direction, or equivalently, the group velocity, is represented by the rate of change of the potential in time, allowing us to fine-tune the simulation to match the physical scenario closely.
 
In this system, the paraxial motion through the propagating channel periodically alters the transverse potential  harmonic force constant,  in the manner of a parametrically driven oscillator.  Thus the Mathieu equation and its attendant stability approximately applies\cite{doi:10.1073/pnas.2110285118}. The actual phase space diagram  for the chosen energy  resembles that of a kicked rotor. In (figure \ref{fig3}), we see the stable island, but at other energies the kicking frequency changes and can become resonant with the transverse mode frequency, destabilizing the motion down the channel. 

The results in Figure~\ref{fig1}D demonstrates that the superwire dynamics, including the tunneling seen in panel A, are successfully (and nearly indistinguishably) captured within the paraxial approximation. This justifies the use of a quasi-one dimensional interpretation when we discuss tunneling in the two-dimensional potential of  Figure~\ref{fig1}A as follows: 

In Figure~\ref{fig1}B(i),  $\Re(\psi)$ displays the consequence of the wavepacket being launched slightly off-center in the channel. This offset creates a combination of the ground and first excited transverse modes, causing a gentle oscillation between the upper and lower bumps as it travels down the channel. 

\begin{figure}[h!]
    \centering
    \includegraphics[width=0.95\linewidth]{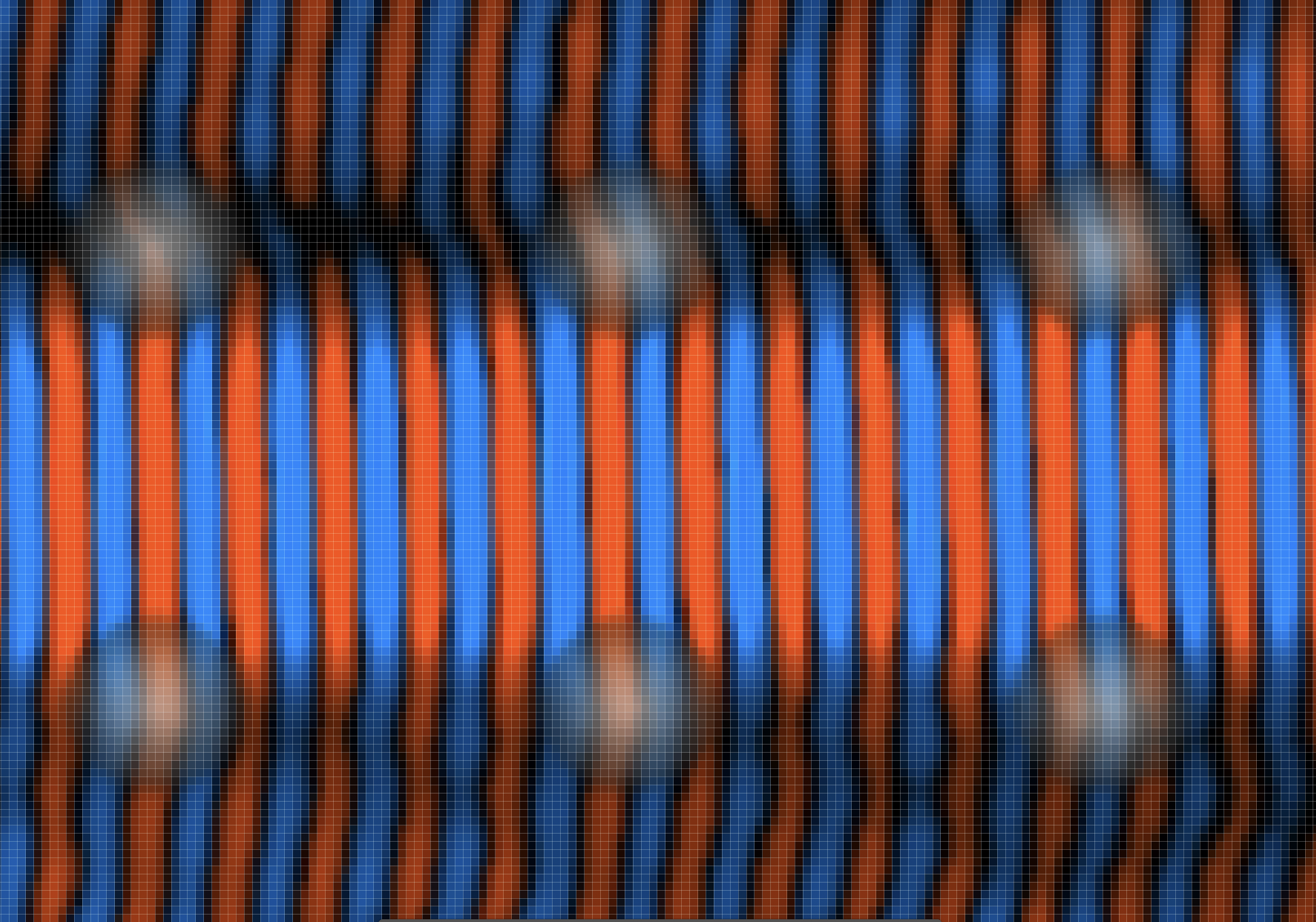}
\caption{Zoomed-in view depicting quantum dynamics featuring dynamical tunneling. The image illustrates the interface between the primary superwire channel and its adjacent channels. Transparent features, depicted in grayscale, represent the superlattice bumps. Notably, the wave propagates without decay  across the top of these bumps, indicating the are classically surmountable. Examination of the real part of \(\psi\) reveals an approximate phase shift of \(\pi/2\) between each channel.\label{fig_scat}}
\end{figure}

\begin{figure*}[ht]
\includegraphics[width=1\linewidth]{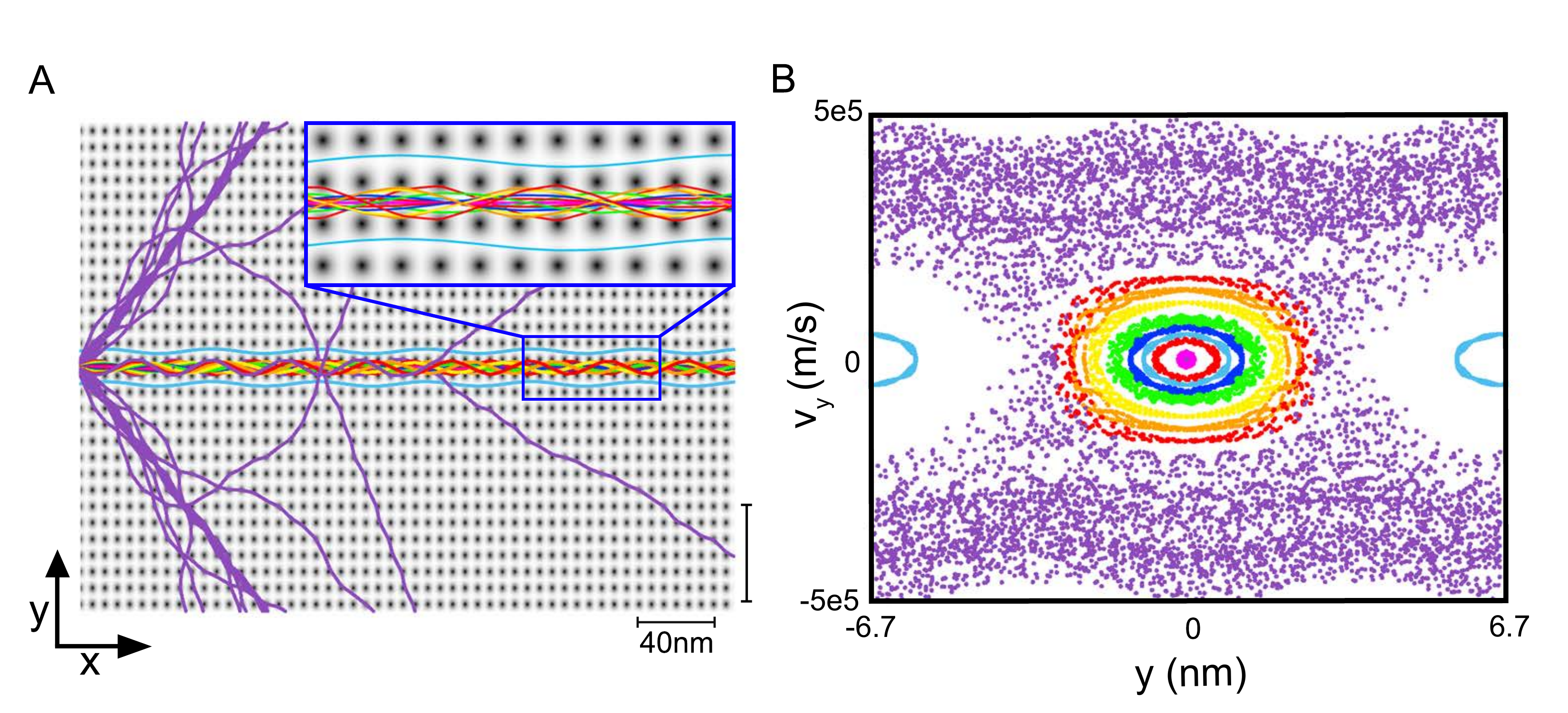}
\caption{ Classical dynamics of superwires in periodic lattices. A) Trajectories of classical particles in coordinate space of x and y. All particles were launched with fixed initial momentum and varying launch angle. Light blue trajectories are translated by a lattice constant to form stable superwire in adjacent horizontal channels. Purple trajectories are not stable inside the horizontal superwire. B) Poincaré surface of section (PSS) for a unit cell with colors corresponding to the trajectories on the left. Stable superwires are indicated by quasi-periodic orbits. The light blue disjoint orbits correspond to disconnected superwires in neighboring channels. Purple dots form the chaotic sea.  \label{fig3}}
\end{figure*}

In this non-integrable superlattice potential, the pseudomomentum components $k_x$ and $k_y$ can  exchange energy in principle, leading to the electron scattering more widely into the 2D lattice. Such high angle scattering  is nonetheless absent here: the stable dynamics  maintains  superwire flow over long distances without significant dispersion, apart from a dynamical tunneling populating the adjacent channels as shown in \ref{fig1}B(ii) and (iii). Figure~\ref{fig_scat} provides a close up of a wave in the middle channel tunneling into  adjacent channels.
The tunneled components continue to propagate in the original direction and the confinement within the next channel is maintained. Remarkably, the oscillatory rocking motion within the first adjacent channels is more pronounced, occurring synchronously on both sides. We do not observe back scattering but we see a phase shift of approximately \(\pi/2\)  between $\Re(\psi)$ in the center channel and those in adjacent channels. Figure~\ref{fig_scat} illustrates that the wave also over rides the potential bumps. The potential barriers are comparable in size to $\lambda_e$ . The relatively low potential strength, substantially reduces the probability of notable diffracting phenomena.

Applying the insight from the paraxial approximation in Figure~\ref{fig1}C(i) helps to understand the underlying tunneling mechanism. The y-component of the superwire eigenstate inside a channel is \textit{closely resembling the 1D eigenstates of a single harmonic quantum well}. The real part of the wave function, $\Re(\psi(y,t))$, demonstrates that the center channel is made of linear combinations of ground state-like (abbreviated GS; Gaussian without a node) and first excited state-like (abbreviated 1ES; Gaussian with one node) components. The composition of the states in each "well" is schematically depicted at the bottom of the figure: Large oscillator symbols indicate major contribution, small symbols a minor, and missing symbols negligible contribution to the wavefunction.

In panel C(i) the absolute value reflects this as a slightly deformed Gaussian. The tunneled components in the adjacent channels are still relatively small but close inspection shows that they carry a node in the middle. Our interpretation is that the adjacent channels at this distance contain a quantum state that is still a mix of GS and 1ES - but the fraction of the component resembling 1ES is already larger here. Hence $\Re(\psi)$ and the absolute value show a node. This is expected because, despite the different mechanisms between barrier and dynamical tunneling, modes with higher energy typically tunnel faster. This behaviour will become more pronounced in the following zoomed in sections C(ii) and C(iii) and supports the claim that dynamical tunneling in superwires could be used to separate the different transverse  energy components of the channel eigenstates. 


There are several indications in Figure~\ref{fig1} of more facile tunneling of the first excited transverse mode of the central channel: damping of the central channel oscillations with progress to the right, and a higher population of 1ES over GS in the side channels compared to the parent channel as further shown in the later sections of the superwire.

\section{Insights from Classical Superwires} \label{Section3}

Tunneling is a quantum mechanical phenomenon that allows particles to transition between states or regions that are not directly connected or accessible according to classical mechanics but they do not have to involve a potential barrier, as is the case here. Dynamical tunneling occurs between classically disjoint regions in phase space. We will also see that here the classically distinct regions are separated by a chaotic sea, so the tunneling falls into another class: the chaos-assisted tunneling of Tomsovic and Ullmo\cite{PhysRevE.50.145}. 
Classical particles are initialized with defined velocity $v$ (corresponding to the expectation value in Section~\ref{Section2}) and positioned at the bottom of the center channel. By adjusting their initial angles, we explore a range of trajectories shown in Figure~\ref{fig2}A. Utilizing a Verlet integration scheme, we calculate the particle motion over time, enabling us to map their trajectories and analyze the associated phase space. To simplify the analysis of the four-dimensional phase space, we employ Poincaré surfaces of section (PSS), focusing on slices along the $v_y$ and $y$ dimensions, as depicted in Figure~\ref{fig2}B. The position of each particle, recorded at intervals equal to integer multiples of the lattice constant from the starting point, is projected onto this PSS, facilitating a comprehensive phase space examination projected on the unit cell of the square lattice.

Figure~\ref{fig2} demonstrates clearly that most trajectories in this parameter regime (pink, blue, red, green, yellow, orange and red) form stable wires inside the center channel, maintaining their confinement over extended distances. This stability is primarily due to what is termed the "rocking-chair" motion: trajectories oscillate in a manner that is slightly off-resonance with the lattice's periodic potential, preventing escape. This off-resonant oscillation underpinning the dynamical self-confinement of these superwire trajectories, is corroborated by the emergence of stable (so-called) calm islands in the PSS shown in Figure \ref{fig2}B, where quasi-periodic elliptical orbits manifest. According to predictions by the Kolmogorov–Arnold–Moser (KAM) theorem, quasi-periodic motions persists within these elliptically bounded islands. 

Each colored elliptical region matches the trajectory color in the panel to the left. Remarkably, despite interaction with a non-integrable potential, these trajectories exhibit almost periodic motion in the PSS.
In contrast, classical particles wandering through a periodic lattice but not confined by channels, typically are chaotic, and have been shown to exhibits features, such as branched flow~\cite{heller2021branched}  and Lévy flights~\cite{lorentzgas}.

The light blue trajectories correspond to the same initial conditions but launched in the neighboring channel. Upon examining the light blue orbits in the PSS, it is evident they occupy separate elliptical orbits, completely disjoint in phase space. This observation implies that, from a classical perspective, a superwire trajectory remains dynamically confined, preventing it from migrating into adjacent channels whereas quantum mechanically we find probability density to tunnel across the disjoint regions in phase space indicated by the formation of the parallel wires in neighboring channels.

The trajectories shown in light blue originate from identical initial conditions but are initiated in neighboring channels. When analyzing these light blue paths on the PSS, it's clear they follow distinct elliptical orbits that do not overlap in phase space. This separation suggests that, according to classical mechanics, a trajectory within a superwire is dynamically confined, which inhibits its movement into adjacent channels. In contrast, the quantum mechanical simulations show that  probability density tunnels through these disjoint regions in phase space, as demonstrated by the presence of parallel superwires in neighboring channels.


We also plot (purple) unstable classical trajectories that do not remain confined within the horizontal superwire. Figure \ref{fig2}A demonstrates that trajectories tending towards instability are typically associated with larger launch angles. Some of these trajectories find channels at different angles, occasionally forming less populated wires. A few paths initially follow the horizontal channel but fail to meet the established criteria for stability, diverging at greater distances. In this regime, certain particles are deflected, leading to early signs of branched flow. Crucially, these diverging trajectories cross the potential peaks, showing that they are not restricted by any potential energy confinement. These unstable trajectories contribute to the scattered data points in the PSS, together creating the ``chaotic sea''—a region in phase space where dynamics become chaotic. The stable quasi-periodic orbits within and adjacent to the central channel are distinctly separated by the chaotic sea. Therefore, particle dynamics within the periodic lattice precludes transitioning between superwire channels without encountering this chaotic region. Thus, the shift from one superwire to another is classically forbidden, indicating that observed transitions in quantum superwires are indeed the result of dynamical tunneling and more specifically chaos-assisted tunneling. 

\section{Controlling Dynamical Tunneling in Engineered Superlattice Structures} \label{Section4}

Although superwire stability is generally resilient to minor parameter changes, the tunneling rate into adjacent channels is highly sensitive to these variations. Given the vast array of possible lattice configurations, a comprehensive study of each is impractical. Nonetheless, we demonstrate that the tunneling rate can be effectively controlled through minor modifications in the superlattice fabrication process.

The tunneling is influenced by the parameter $\alpha$ in the Fermi function of the superlattice, which affects both the sharpness of convex features on lattice sites and their maximum height. Figure \ref{fig4} provides a detailed view of the superlattice, displaying bump heights in grayscale. The inset details the variation in bump height with distance from lattice points for three distinct $\alpha$ values.

\begin{figure*}[ht]
\includegraphics[width=\linewidth]{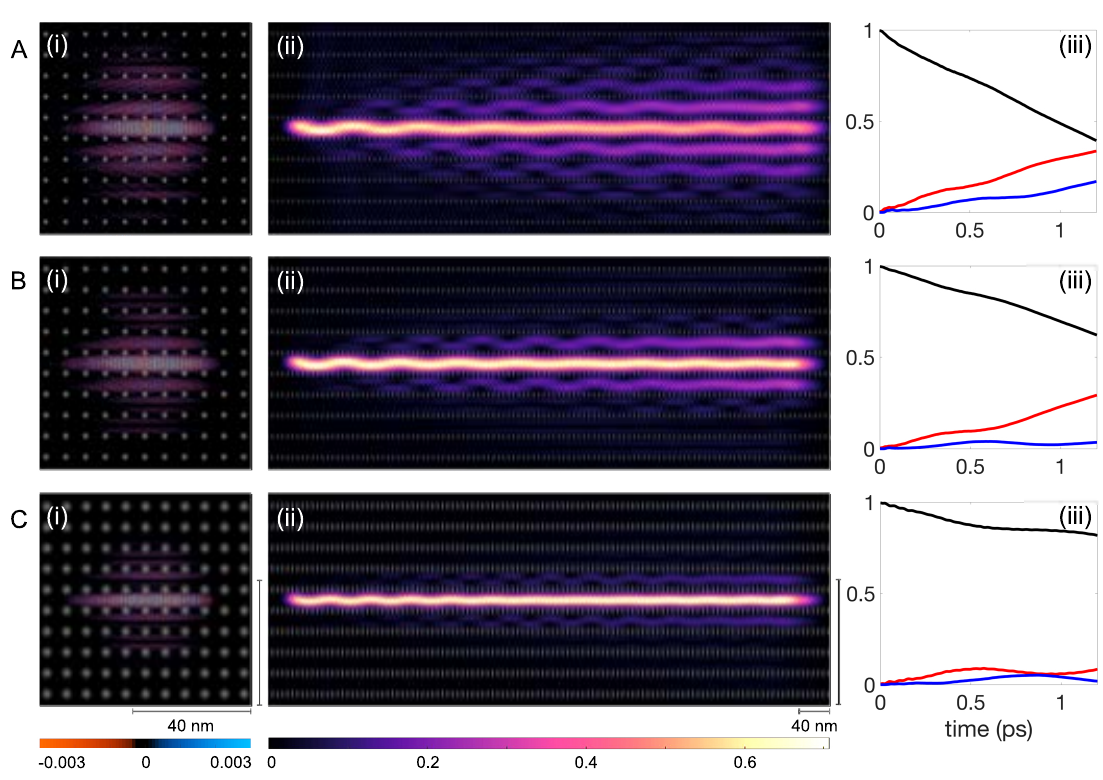}
\caption{Study of tunneling rates with varying lattice features. A), B), and C) show the results for the different $\alpha$ values at 0.03, 0.05 and 0.10, respectively. For each case panels (i) corresponds to real part of the wavefunction $\psi(x,y,t)$ at $t=0.66$ ps. Panels (ii) correspond to the probability densities $P(x,t)$ of the eigenstates in each scenario. Panels (iii) show the probability densities in the center channel ($P_0$, black line), the first adjacent channels combined ($P_1$, red line) and the second adjacent channels combined ($P_2$, blue line) for all $\alpha$s used here. Note that the wave oscillations in the channel diminish from left to right. as the first transverse excited state preferentially escapes the center channel, leaving it a higher percentage of  ground state amplitude heading from left to right. In panel A(ii) especially, it can be seen that channels furthest from the initially populated center channel become increasingly purely the first excited transverse mode.}\label{fig5}
\end{figure*}

We conducted wavepacket dynamics simulations consistent with those described in Section \ref{Section2}. Figure \ref{fig5} displays snapshots of $\Re(\psi)$ at a specific timestep for three superlattice configurations in the left column: panel A(i) for $\alpha$ = 0.03, panel B(i) for $\alpha$ = 0.0, and panel C(i) for $\alpha$ = 0.10. The snapshots reveal that smoother potential features, despite a slight reduction in potential height, result in better confinement of $\Re(\psi)$. The images also support that most of $\Re(\psi)$ are contained within 4/5 of the simulation box and the matrix extension technique can be applied.
The probability distribution, $P(x,y)$, of the eigenfunctions for the entire run is presented in the middle column of Figure \ref{fig5} and directly corresponds to the $\Re(\psi)$ images on their left. With increasing $\alpha$, $P(x,y)$ demonstrates improved confinement over longer distances with smoother potential features. For sharp potential features, $\alpha$ = 0.03 in panel A(ii), $P(x,y)$ gets rapidly distributed over adjacent channels, with a notable reduction in intensity within the central channel, indicating a high tunneling rate. Again we can recognize that at the end of the simulation the GS and 1ES components (indicated by nodes or rocking motion in 2D) of the eigenstate get filtered successively with tunneling into each parallel channel. At the end of the simulation the outermost channel always contains the highest 1ES component. For $\alpha$ = 0.05, panel B(ii), the majority of $P(x,y)$ is confined within the center and first adjacent channels on either side. With $\alpha$ = 0.10, panel C(ii), the distribution primarily stays within the center channels and small contributions in the first adjacent channels, demonstrating the effect of $\alpha$ on confinement over experimentally relevant distances.

To quantify the tunneling rates in more detail, we present the distribution of probability density, \(P(x,y)\), across different channels: the center (\(P_0\), black line), the first adjacent (\(P_1\), red line), and the second adjacent (\(P_2\), blue line) in the right column of Figure~\ref{fig5}. For a minimal \(\alpha\) value of 0.03, depicted in panel A(iii), \(P_0\) starts at value one, indicating complete initial confinement within the center channel. The probability \(P_0\) exhibits a nearly linear decline. Concurrently, \(P_1\) and \(P_2\) increase linearly, albeit at a reduced rate, with \(P_2\) lagging slightly until sufficient probability density has transitioned into the first adjacent channels. Remarkably, by the conclusion of the simulation over a span of 700 nm, \(P_0\) and \(P_1\) converge towards comparable values, i.e. the combined probability in the first two adjacent channels is comparable to that of the center channel. In this regime, the tunneling process can be viewed as interaction of an infinite sequence of potential wells. Over time, \(P(x,y)\) disperses uniformly across all channels, akin to behavior in an open system, leading \(P_0\) to approach zero, followed by a similar depletion in the first and second adjacent channels. We call this regime "tunneling escape".

The slight increase in \(\alpha\) to 0.05 impacts the tunneling dynamics quantitatively as demonstrated in Figure \ref{fig5}B. The probability  \(P_0\) still decreases linearly, albeit at a reduced rate, while \(P_1\) exhibits a further delayed linear increase, and \(P_2\) stays near zero, displaying minimal fluctuation throughout the simulation. 

For the highest \(\alpha\) value of 0.10, Figure\ref{fig5}C shows the tunneling dynamics change qualitatively. Initially, \(P_0\) decreases linearly but reaches a temporary plateau, coinciding with the peak and subsequent oscillation of \(P_1\). The probability \(P_2\) remains minimal throughout the simulation but mimics \(P_1\)'s oscillations with a phase difference. This behavior suggests a distinct tunneling regime closer to a three-well problem, characterized by suppressed tunneling rates. Such suppression allows for bidirectional tunneling, contributing to the observed oscillations in tunneling rates. We anticipate that, over extended periods, the superwire would still undergo tunneling escape, but at significantly prolonged timescales, thereby delaying thermalization.
The tunneling rate from the center channel was notably lowest for an \(\alpha\) value of 0.10, despite a marginal decrease in potential maximum relative to other \(\alpha\) values. This \(\alpha\)-dependent behavior underscores that confinement is primarily due to quasi-periodic scattering/deflections of the wavefunction across the potential landscape. Thus, the tunneling mechanism relies not solely on barrier height but is significantly affected by the potential's smooth contours. Furthermore, tunneling can be nearly eliminated by disrupting the overlap between neighboring states, for instance, by creating channels with varying bottom heights or employing nonuniform bump heights in fabrication. These strategies highlight the potential for precise control over dynamical tunneling through tailored potential landscapes. 

We want to point out that the level of (classical) chaos does not change monotonically when the smoothness of the bumps varies from the fully chaotic case of the paradigmatic Lorentz gas composed of hard-wall disk to regular dynamics of the high-softness limit resembling an integrable cosine potential. Instead, there is a rich structure of normal and anomalous diffusion depending on the smoothness (and separation) of the bumps, as reported in Ref.~\cite{lorentzgas} for a similar soft Lorentz gas. An interesting avenue of future research is to investigate how this diversity in classical dynamics affects chaos-assisted dynamical tunneling between channels, alongside with the more general studies on the robustness of superwires, e.g., against various defects and impurities within the lattice.

\section{Conclusion and Future Directions} \label{Section5}

The conduction problem in superlattices thrusts us into the unfamiliar territory of high Brillioun zones and robust efficacy of semiclassical correspondence, since electron wavelengths can become short compared to smooth features of the potential. Here we have used the classical motion to define the existence of tunneling between channels that are otherwise stable quantum mechanically.

This  is the first study of dynamical tunneling in flatband superwires.  Our model system has revealed that even when superlattice features are relatively weak compared to the kinetic energy of the electron, the dynamical confinement within superwires remains robust. This may make it relatively easier to fabricate superwires

Flat bands and channeling superwires go hand in hand. The channel eigenstates  are linear combinations of conventional degenerate Bloch states. This realization opens up new avenues for leveraging superwires in advanced nanodevices, promising novel approaches to electron confinement and transport. 

Through the use of both quantum mechanical and classical methods, we have shown that dynamical tunneling is distinct from traditional barrier tunneling, offering a nuanced understanding of superwire stability and the conditions necessary for their formation and stability. Because of the  chaotic zones intervening between channels, tunneling between them should be further understood as chaos-assisted\cite{PhysRevE.50.145}. 

By use of a paraxial approximation we have been able to understand dynamical tunneling as a quasi-one-dimensional process in a time-dependent potential. This model allows us to approximate the superwires by ground and first excited states in series of periodically driven harmonic quantum wells. 

Analyzing the effects of superlattice parameter variations, particularly the \(\alpha\) parameter, on electron tunneling and confinement, we found pathways for potential control schemes of electron transport in nanoscale devices. For example, we showed how the channel tunneling rate depends on the transverse mode of the channel, thus filtering  energy across multiple parallel channels. This may have implications as a mechanism for energy component filtration and quantum state purification. 

The findings suggest exciting prospects for the design and fabrication of quantum devices, where electron can be guided in quasi one-dimensional channels under consideration of varying superlattice parameters. Current investigations are focused on understanding how multiple electrons interact both within individual channels and across neighboring ones, and on the question of limited pathways for electron-vibration interactions in the channels. The inherent decrease of phonon-electron interactions within these quasi-one-dimensional wires could be an important step to realize near zero-resistivity transport of electrons in superlattices

Exploring the stability of superwires in real-world materials presents a fascinating challenge, especially when considering small lattice displacements or within dynamically disordered potential landscapes, for a classical perspective on this issue, see Ref.~\cite{bernick1995loss}. Our formalism accommodates various approaches to explore both frozen and dynamical deformation potentials in crystal lattices described Ref.~\cite{Plankian_Aydin, DDP_manuscript}.

We hope that future research will explore experimental validation (e.g. through STM studies of twisted multi-layered materials) to further harness superwires and their inherent tunneling for device applications. It is important to note that the principles of superwires and dynamical tunneling extend to photons and quantum-optical systems, suggesting potential experimental investigations in optical setups, including photorefractive crystals and various photonic structures. 

The ability to manipulate tunneling dynamics through tailored potential landscapes offers a pathway to nearly eliminate tunneling, presenting strategies for enhanced electron localization and stability. On the other hand, controlled tunneling can be used to couple initially isolated quantum channels. 

\setcounter{secnumdepth}{0}

\begin{acknowledgments}

We want to thank everyone who provided feedback and support through constructive discussions. Specifically we want to thank Alhun Aydin, Esa R\"as\"anen, Ragnar Fleischmann, Lev Kaplan, Jairo Velasco, Norman Yao, Arthur Jaffe, Roy Garcia and Kaifeng Bu for valuable comments. 
A. M. G. wants to thank the Harvard Quantum Initiative for financial support. M.K. thanks the Harvard College Research Program for financial support. J.K.-R. thanks the Emil Aaltonen Foundation, Vaisala Foundation, and the Oskar Huttunen Foundation for financial support.
\end{acknowledgments}

\section{Appendix} \label{Appendix}

\appendix
\section[\appendixname~\thesection]{}

\subsection[\appendixname~\thesubsection]{Appendix A: Calculation of band structure} \label{A2}

Tight-binding models typically assume electrons are localized at atomic sites with a finite probability of hopping to adjacent sites. However, this framework proves inadequate for electrons in superlattices, where their presence extends between potential peaks. Consequently, we adopt a continuous model approach to accurately capture the electronic band structure in superlattices, grounded in the time-independent Schrödinger equation

\begin{equation}
\left[ -\frac{\hbar^2}{2m} \nabla^2 + V(\vec{r}) \right] \phi(\vec{r}) = E\phi(\vec{r}. 
\end{equation}
According to the Bloch theorem, we have $\phi(\vec{r}) = e^{i\vec{k}\cdot\vec{r}}u_{k}(\vec{r})$, satisfying the following equation.
\begin{equation}
\left[ -\frac{\hbar^2}{2m} \nabla^2 + V(\vec{r}) \right] e^{i\vec{k}\cdot\vec{r}}u_{k}(\vec{r}) = E(\vec{k})e^{i\vec{k}\cdot\vec{r}}u_{k}(\vec{r}).
\end{equation}
In order to reduce our calculation, we only focus on the periodic potential in a unit cell. By multiplying \( e^{-i\vec{k}\cdot\vec{r}} \) on both sides, the equation reduces to

\begin{equation}
\begin{split}
e^{-i\vec{k}\cdot\vec{r}}\left[ -\frac{\hbar^2}{2m} \nabla^2 + V(\vec{r}) \right] e^{i\vec{k}\cdot\vec{r}}u_{k}(\vec{r}) \\
= e^{-i\vec{k}\cdot\vec{r}}E(\vec{k})e^{i\vec{k}\cdot\vec{r}}u_{k}(\vec{r}).
\end{split}
\end{equation}
Since \( e^{-i\vec{k}\cdot\vec{r}} \) and \( e^{i\vec{k}\cdot\vec{r}} \) cancel each other on the right hand side, we obtain the following equation,
\begin{equation}
\begin{aligned}
H_{k} u_{k}(\vec{r}) = \left[ \frac{\hbar^2}{2m}(-i\nabla + \vec{k})^2 + V(\vec{r}) \right] u_{k}(\vec{r}) = E u_{k}(\vec{r}) \quad 
\end{aligned}
\end{equation}
where \( E(\vec{k}) \) is the eigenvalue and \( u_{k}(\vec{r}) \) is the Bloch state.
For a two-dimensional superlattice, we get
\begin{equation}
\begin{split}
H_k &= \bigg[ \frac{\hbar^2}{2m}(-i\nabla + \vec{k})^2 + V(\vec{r}) \bigg] \quad \\
=&\Bigg[ \frac{\hbar^2}{2m}\bigg(-\frac{d^2}{dx^2}-\frac{d^2}{dy^2} - ik_x\frac{d}{dx}\\ &-ik_y\frac{d}{dy}+ k^2\bigg) + V(x) \Bigg].
\end{split}
\end{equation}

Given the exact form of potential in a unit cell, diagonalization process of Hamiltonian can be executed by constructing the numerical form of the $\nabla$ operator. Thus, we can get the full 2D electronic band structure.

\subsection[\appendixname~\thesubsection]{Appendix B: Wavepacket dynamics with matrix extension} \label{A1}

We are employing the well-known split-operator method (with Strang splitting) to propagate the wavefunction in time with the kinetic energy operator $T$ and the potential energy operator $V$. Each iteration is using the following steps: 

1. Apply the potential term in real space for one half-time step:

\begin{equation}
\psi(x, t + \Delta t / 2) = e^{-i V(x) \Delta t / 2} \psi(x, t)
\end{equation}

2. Fourier transform to momentum space:

\begin{equation}
\tilde{\psi}(k, t + \Delta t / 2) = \mathcal{F}\{\psi(x, t + \Delta t / 2)\}
\end{equation}

3. Apply the kinetic term in momentum space:

\begin{equation}
\tilde{\psi}(k, t + \Delta t) = e^{-i T(k) \Delta t} \tilde{\psi}(k, t + \Delta t / 2),
\end{equation}

where \(T(k) = \frac{\hbar^2 k^2}{2m}\) is the kinetic energy operator in momentum space, with \(k\) being the wavevector, \(\hbar\) the reduced Planck's constant, and \(m\) the mass of the particle.

4. Inverse Fourier transform to return to real space:

\begin{equation}
\psi(x, t + \Delta t) = \mathcal{F}^{-1}\{\tilde{\psi}(k, t + \Delta t)\}
\end{equation}

5. Apply the potential term again in real space for another half-time step:

\begin{equation}
\psi(x, t + \Delta t) = e^{-i V(x) \Delta t / 2} \psi(x, t + \Delta t)
\end{equation}

\noindent
Together, the full step from \(t\) to \(t + \Delta t\) can be written as:

\begin{equation}
\begin{split}
\psi(x, t + \Delta t) &=
e^{-i V(x) \Delta t / 2} \mathcal{F}^{-1}\{e^{-i T(k) \Delta t}\\ &\times\mathcal{F}\{e^{-i V(x) \Delta t / 2} \psi(x, t)\}\}
\end{split}
\end{equation}

With this algorithm applied to superwires the wavefunction predominantly propagates confined within a channel in +x direction (almost no backscattering). This special case of a quasi 1D localization allows to follow the wavefunction over multiple simulation boxes without changing the size of the grid at any given time step. Instead we move the frame as follows: 

When some small probability, defined by an adjustable threshold value, is found in the right-most column of the position grid, we shift our frame to the right, following the main part of the wavefunction. This is specifically achieved by leveraging the potential's symmetry, wherein we augment the simulation grid by appending a potential matrix one-fifth (equivalent to three unit cells) the width of the position grid. This extension provides additional space, allowing the wavefunction to continue its propagation towards the right. To preserve the original grid dimensions, we remove the initial one-fifth of the simulation box, a section that, upon careful examination, contains less than 1\% of the probability density. 
In our simulations, we ensure that forward scattering predominates, preventing the wavepacket from backscattering and self-interference. As a result, the eigenstates are extractable from the wavepacket after a single traverse through the simulation box. Subsequent to each matrix extension step, the ensuing segment of the eigenstate is spatially appended to the preceding frame, resulting in a composite eigenstate seamlessly spanning the entire distance it has traversed.
It is important to note that the matrix extension technique can potentially lead to amplitude truncation if the wavepacket extends excessively into the matrix section slated for removal during the extension process. To mitigate this, we calibrated the simulation parameters to ensure that, within the simulation duration, over 99\% of the wavepacket's probability density remains outside the excised region. The eliminated components are primarily minor amplitude constituents that propagate slower or are directionally shed from the initial trajectory at launch, which may be regarded as initial impurities within simulation. Minimal leakage towards the top and bottom of the frame is effectively mitigated by implementing a smooth absorption boundary.


\bibliography{references}

\end{document}